# Central symmetry and antisymmetry of the microwave background inhomogeneities on Wilkinson Microwave Anisotropy Probe maps

Iurii   Kudriavtcev

Dmitry A. Semenov

We performed a visual and  numeric analysis of the deviation of the microwave background temperature on WMAP maps. We proved that the microwave background inhomogeneities possess the property of the central symmetry resulting from the two kinds of central symmetry of the opposite signs.

After the computer modeling we have established the relation between the coefficient of the central symmetry and the values of the symmetrical and antisymmetrical components of the deviation of the temperature. The obtained distribution of the symmetry coefficient on the map of the celestial sphere in Mollweide projection testifies on a contribution of both kinds of central symmetry which is approximately equal on the average in absolute magnitude but opposite by sign and where one kind of the central symmetry prevails on some sections of the celestial sphere and another kind – on the others. The average resulting value of the symmetry coefficient on the sections with angular measures less than $15\text{-}20^0$ varies within the range from -50% to +50% with some prevalence of the antisymmetry – the average coefficient of the central symmetry for the whole celestial sphere is $-4\pm1\%$. (antisymmetry 4%).

Small scale structure of the distribution indicates that it is the result of the combined action of the mechanisms of the central symmetry and central antisymmetry, close to 100%.

98.80.-k

## 1. Introduction

The symmetry of the large scale microwave background inhomogeneities that was recently discovered at the WMAP maps analysis [1]  is interpreted as the axisymmetry of the Universe, or in other words, as the existence of the distinguished symmetry axis in the space which contradicts to the isotropy principle of the space that is fundamental for the Relativity Theory. In fact, the image obtained in [1] which is the image of the distribution of the large scale inhomogeneities on the celestial sphere in Mollweide projection (picture from the web site prl.aps.org), presented on the Picture 1, looks axisymmetrical.

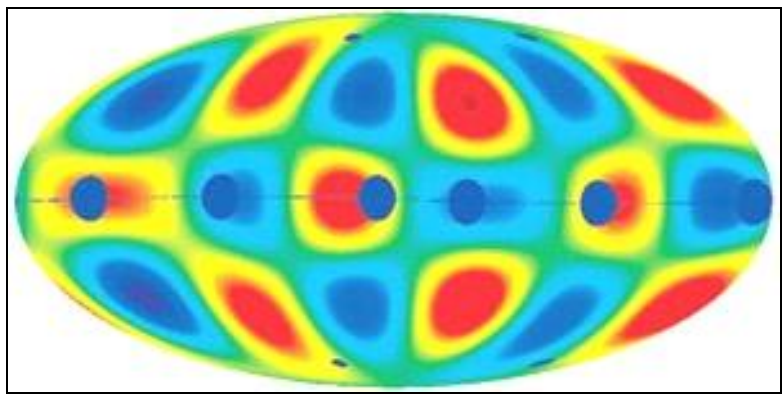

**Picture 1.**  Large scale symmetry of the microwave background inhomogeneities on the celestial sphere (image from the web site prl.aps.org, presented in gray color scheme). The maximums of the positive deviations of the temperature (red on the original picture) look darker, the ones of the negative ones (blue on the original picture) look lighter.

But another interpretation of this distribution is also possible, the one which stays within the basics of the Relativity Theory. The particularity of the microwave background shown on the Picture 1 can be interpreted not only as an axisymmetry but also as a central symmetry, or more precisely, central antisymmetry, when each inhomogeneity of the microwave background temperature corresponds to the same inhomogeneity of the opposite sign in the opposite, centrally symmetrical point of the celestial sphere. Such interpretation does not require a distinguished axis which means it does not contradict to the space isotropy requirement but it also does not fully correspond to the standard cosmological model as its realization is possible only when the dynamics of the Universe development are slower than the ones of the standard model and when the light waves, disengaged at the moment of recombination and having the information about the corresponding distribution of the inhomogeneities move to us longer and cover the bigger distances than the ones in the standard model, meaning being bigger that the length of the half-circle of the closed Universe.

As the violation of the basic isotropy principle, fundamental for the Relativity Theory is far more important than the contradiction to the requirements of its particular applications including the cosmological models, we consider it reasonable to confirm the hypothesis of the interpretation of the microwave background inhomogeneities from the point of view of the central symmetry, regardless its correspondence to the cosmological models accepted at the moment. At that it is important to keep only one condition – the closed model of the Universe that describes the expansion of the 3-dimensional hypersphere in a 4-dimensional space [2].

Let us see the light radiated by the inhomogeneity in a certain spot of the hypersphere in the period of recombination. Let us assume that the observer stands in another spot and see this signal after several rotations of the ray around the hypersphere. It is not difficult to see that despite the fact that the inhomogeneity radiated the light in all the directions we received only the light that was spread in our direction by the arc of the big circle (if we use the terms of the 3-dimensional model). At that however many full rotations this light could make by the arc of the big circle it will be getting back to us again and again. That is why it is not important to us how many rotations it made until we see it in our telescope. What is important is that the light from the same inhomogeneity will always come to us by two ways that correspond to the parts of the big circle.

If in some point of the celestial sphere we see the microwave background inhomogeneity that is induced by the movement of the matter at the moment of recombination and corresponds to the positive deviation of the temperature (the velocity of the radiating matter is directed to us – it approaches by this arc) then at the same time we should see the same inhomogeneity in the opposite point of the celestial sphere (signal that came by another arc of the big circle, or "from behind") but with the negative deviation of the temperature as it leaves by this arc. That is what we see on the Picture 1 – the positive maximums coincide with the opposed negative ones. Let us note that for the inhomogeneities that are induced by not the movement of the matter but by its condensation, it is natural to assume the existence of the central symmetry with not a positive but a negative sign. This way in the described model we can expect a simultaneous appearance of both kinds of central symmetry on the celestial sphere.

**2. Visual search of the elements of the central symmetry on WPAM maps in Lambert and Mollweide projections.**

Let us examine if these effects of the central antisymmetry appear not only on the large scale microwave background inhomogeneities as it is shown on the Picture 1 but also on the smaller ones. Initially the presence or the absence of the searched effect can be checked by means of the visual analysis of the images we have. The most comfortable for this purpose seem to be the image of the microwave background inhomogeneities on WMAP maps in azimuthal equal-area Lambert projection [3], that are represented by a color image of the deviations of the microwave background temperature.

To get a pair of comparable images the identical (identically spaced) points of which correspond to the diametrically opposite points of the celestial sphere, the initial image of one of the hemispheres (right one, in our case) should be inverted. To reveal the antisymmetry effect we deprive the image of colors that correspond to the deviations of the temperature of the opposite sign on the opposite hemispheres. After this processing we obtain the pairs of the images of the hemispheres as it is shown on the Picture 2.

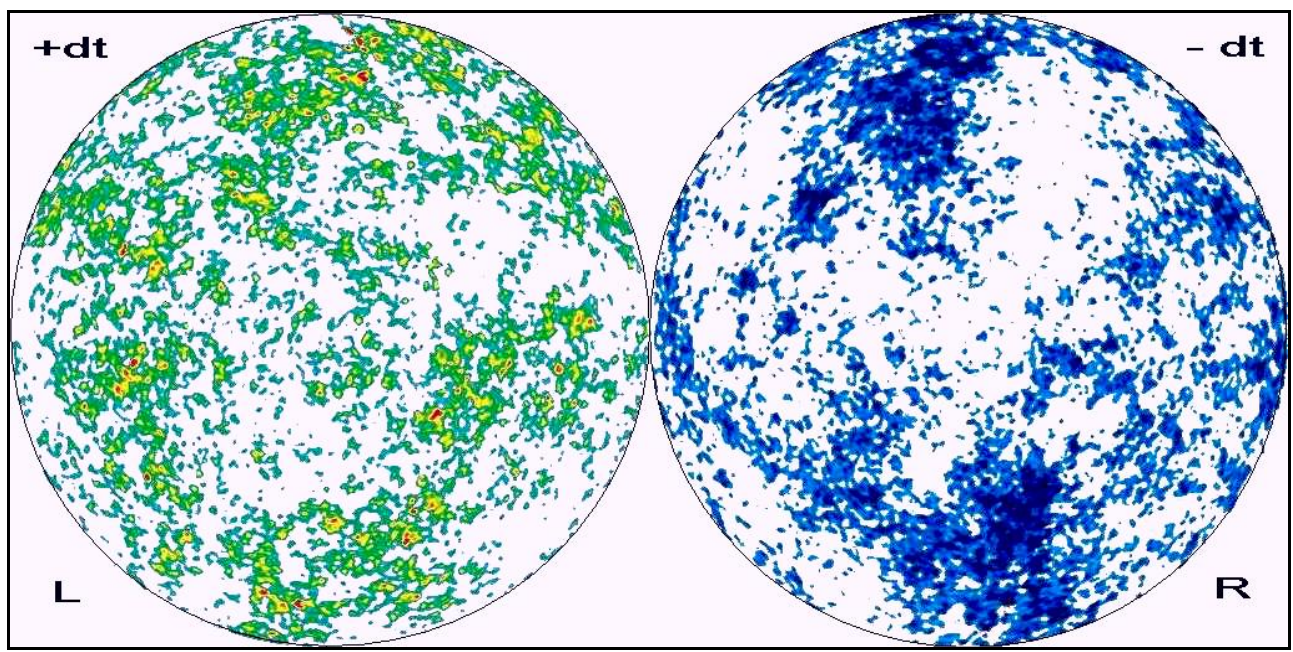

**Picture 2.** Two images of celestial hemispheres obtained as a result of processing of the initial WMAP pictures. On the left hemisphere we kept the colors that correspond to the positive deviations of the temperature, on the left – to the negative ones. It is shown in a gray color scheme. The picture of the right hemisphere is inverted.

The identically spaced points on these pictures correspond to the diametrically opposite points of the celestial sphere. Visual comparative analysis of this pair of images allows to notice the existence of a large scale correlation of the inhomogeneities of the opposite sign which corresponds in general to the results presented on Picture 1. At that we can point out the similarity of smaller details of the images as well but we do not observe strong correlation of them.

For a more confident visual revelation of the coinciding elements we can use the ability of the visual analyzer to reveal the mirror symmetry of the studied images. With this purpose we selected and combines in pairs the identically spaced fragments of the left and right images where the fragments of the right image are taken in mirror reflection. On the Picture we present an example of such a pair of the fragments of the images from Picture 2. We should point out that the left and the right halves of the presented pair of fragments belong to the opposite celestial hemispheres and are placed there centrally symmetrically.

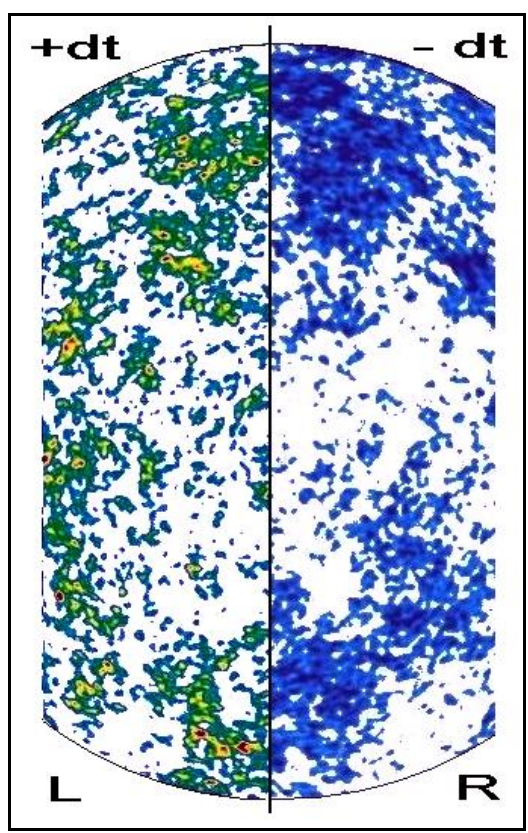

**Picture 3.** Comparison of the picture fragments of the microwave background inhomogeneities that correspond to the diametrically opposite segments of the celestial sphere. Right halves of the fragments is taken in mirror reflection.

On the pair of fragments presented on the Picture we can clearly see the elements of the mirror symmetry both large and some small elements of the images which confirms the fact of the existence of the central antisymmetry for the large scale and smaller scale microwave background inhomogeneities. At the same time the picture of antisymmetry is severely distorted and darkened in many places by non-symmetrical inhomogeneities which allows to assume the existence of several simultaneously working mechanisms of microwave background formation. The same picture can be observed on the other analyzed pairs of fragments of celestial hemispheres (see Appendix). We can particularly note the elements of the images that can be interpreted as a result of the action of the second mechanism mentioned above – central symmetry of positive sign.

The same visual analysis can be performed on the WMAP map in Mollweide projection as well [3], where we can combine without significant distortion the centrally symmetrical segments in the equatorial zone by the shifting to the half length of the equator ($180^0$). On the Picture 4 we present the pair of fragments obtained as the result of such shifting of the subequatorial zone of the width of 10% of the map size which makes the angular measure about $15^0$.

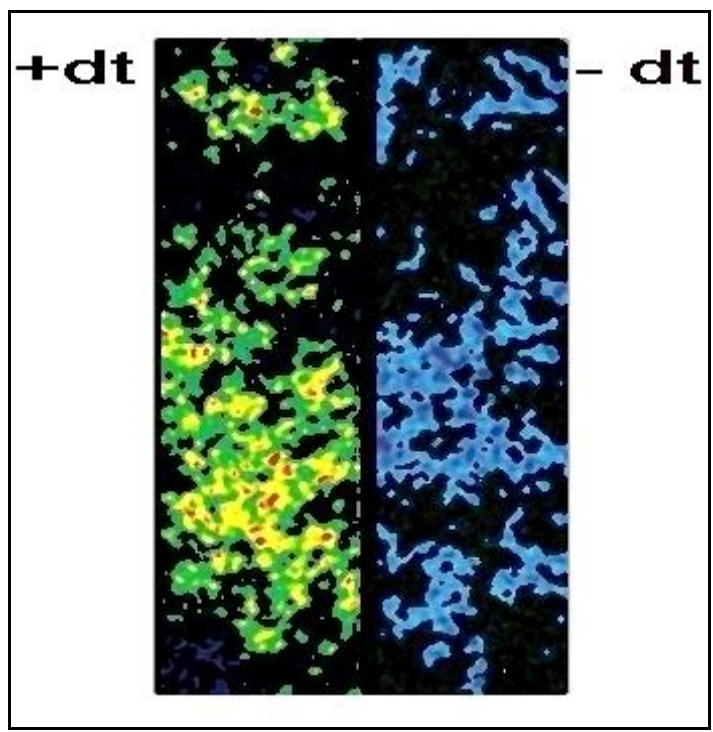

**Picture 4.** Manifestation of the central antisymmetry of the microwave background on the WMAP map in Mollweide projection. We compare a pair of fragments of the subequatorial zone with angular width about $15^0$, shifted relatively to each other by half length of the equator ($180^0$). Presented on the black background, vertically.

The similarity of the centrally symmetrical segments of the map at comparison of the two images shifted relatively to each other by the half length of the equator should be expressed in the appearance of the mirror symmetry elements which is clearly seen on the presented pair of fragments (placed vertically to make visual analysis of the mirror symmetry more comfortable). This way the existence of the central antisymmetry of the microwave background is confirmed on the WMAP map in Mollweide projection. We should notice that the precision of the visual analysis of the symmetry elements depends on the choice of the limits of the cut out segments of the spectrum.

The results of the visual analysis of the central antisymmetry of the microwave background testify on the appropriateness of a more detailed numerical analysis of this phenomenon.

**3. Numerical analysis of the deviations of the microwave background temperature and computer modeling of the central symmetry.**

For the numerical analysis of the deviation of the microwave background temperature we used the data in FITS format (wmap_ilc_7yr_v4.fits) from the web site [3].

The analysis was performed on the assumption that any variable on the spherical surface can be presented as a sum of the symmetrical and antisymmetrical components that are the half sum and the half difference of its values in diametrically opposite points. If $T_1$ и $T_2$ are the deviations of the microwave background temperature in the studied and in the opposite (centrally symmetrical) points of the celestial sphere, then

$$T_{symm} = (T_1 + T_2)/2; \qquad (1)$$
$$T_{asymm} = (T_1 - T_2)/2; \qquad (2)$$

As both these values as well as the initial distribution of the deviation of the temperature T, have a great random ("noise") component and are able to take both positive and negative significances, it appeared to be convenient to analyze their average absolute values (modules) or root mean square. Table 1 contains these values averaged for the whole celestial sphere.

| # | Value | Symbol | Arithmetic average for the whole celestial sphere $T^a(\mu K)$ | Root mean square for the whole celestial sphere $T^q(\mu K)$ |
|---|---|---|---|---|
| 1. | Initial deviation of the temperature | T | -0.77 | 70.57 |
| 2. | Symmetrical component | $T_{symm}$ | -0.76 | 47.88 |
| 3. | Antisymmetrical componen | $T_{asymm}$ | -0.00 | 51.83 |
| 4. | Absolute value of the deviation of the temperature | \|T\| | 55.86 | - |
| 5. | Absolute value of the symmetrical component | \|$T_{symm}$\| | 38.18 | - |
| 6. | Absolute value of the antisymmetrical component | \|$T_{asymm}$\| | 41.21 | - |

**Table 1.** Calculated values of the deviation of the microwave background temperature according to WMAP (wmap_ilc_7yr_v4.fits) and its symmetrical and antisymmetrical components averaged for the whole celestial sphere.

To evaluate the influence of the symmetry on the correlation of the specified average parameters we have done a computer modeling using the random number generator in the Excel tables format. We analyzed a group of 10000 pairs of random numbers X1, X2. Every pair was used to model the values of the deviations of the temperature in the studied and in the opposite to it points of the celestial sphere (T1, T2). Using the expressions (1),(2) we calculated for them the symmetrical and antisymmetrical components, the absolute values, the average arithmetic values and root mean squares for the number group.

The parameters of the noise signal were chosen the way that could allow the approximate correlation of the root mean square of the deviation of the temperature $T^q$ and the average arithmetic of its absolute value $|T^a|$ to the data obtained for the celestial sphere ($|T^a|$ = 55.9 μK, $T^q$ = 70.6 μK, see Table 1). The correlation of these values depend on the exponent at the given by the random number generator random number rand(0,1), uniformly distributed within the range (0,1). The best correlation was reached at X = 142 R$|R|^{0.53}$, where R = 2(rand-0,5). The model values obtained at that (with no symmetry) are presented in the Table 2.

| Value | $\|T^a\|$ (μK) | $\|T^a_{symm}\|$ (μK) | $\|T^a_{asymm}\|$ (μK) | $T^q$ (μK) | $T^q_{symm}$ (μK) | $T^q_{asymm}$ (μK) |
|---|---|---|---|---|---|---|
| Average value of the model from 10000 pairs by 10 attempts | 56.01 | 40.01 | 40.35 | 70.40 | 49.56 | 49.98 |
| Standard deviation of the model from 10000 pairs by 10 attempts | 0.30 | 0.22 | 0.30 | 0.31 | 0.25 | 0.30 |
| Calculated average values for the whole celestial sphere according to WMAP (from Table 1) | 55.86 | 38.18 | 41.21 | 70.57 | 47.88 | 51.83 |

**Table 2.** Average values of the deviation of the microwave background temperature and its symmetrical and antisymmentrical components given by the computer model in comparison with the WMAP values, averaged for the whole celestial sphere.

Symmetry was introduced to the model by adding to each of the number of the pair a certain part of the second number which is given by the coefficient $k_{symm}$. At that the values $T_1$ and $T_2$ are defined by the expressions

$$T_1 = X_1 + k_{symm} X_2; \qquad (3)$$

$$T_2 = X_2 + k_{symm} X_1. \qquad (4)$$

The value $k_{symm} = 0$ corresponds to the absence of the symmetry, $k_{symm} > 0$ - to the positive central symmetry, $k_{symm} < 0$ - to the negative central symmetry (antisymmetry).

The nature of changes of the average values of T, $T_{symm}$, $T_{asymm}$, obtained as a result of modeling when changing the symmetry coefficient within the range ( -0.3 < $k_{symm}$ < 0.3), is presented in the Table 3. There are the average values after 10 calculations on the multitude of 10000 pairs of random numbers for every value of $k_{symm}$. Standard deviations from these values are within the range 0.2 - 0.5 μK, in average 0.31 μK.

| $k_{symm}$ | $\|T^a\|$ (μK) | $\|T^a_{symm}\|$ (μK) | $\|T^a_{asymm}\|$ (μK) | $T^q$ (μK) | $T^q_{symm}$ (μK) | $T^q_{asymm}$ (μK) |
|---|---|---|---|---|---|---|
| -0.30 | 56.01 | 28.06 | 52.37 | 70.40 | 34.79 | 64.93 |
| -0.20 | 56.33 | 32.33 | 48.24 | 70.66 | 40.02 | 59.76 |
| -0.10 | 56.38 | 36.24 | 44.33 | 70.71 | 44.88 | 54.88 |
| -0.08 | 56.18 | 36.95 | 43.62 | 70.56 | 45.79 | 53.97 |
| -0.06 | 56.03 | 37.93 | 42.56 | 70.43 | 46.92 | 52.66 |
| -0.04 | 56.08 | 38.61 | 41.90 | 70.47 | 47.76 | 51.88 |
| -0.02 | 56.03 | 39.29 | 41.11 | 70.41 | 48.77 | 50.96 |
| **0.00** | **56.01** | **40.01** | **40.35** | **70.40** | **49.56** | **49.98** |
| 0.02 | 56.09 | 40.84 | 39.45 | 70.44 | 50.65 | 48.84 |
| 0.04 | 55.97 | 41.83 | 38.62 | 70.37 | 51.78 | 47.88 |
| 0.06 | 56.17 | 42.60 | 37.83 | 70.52 | 52.78 | 46.90 |
| 0.08 | 56.21 | 43.48 | 37.12 | 70.54 | 53.88 | 45.99 |
| 0.10 | 56.12 | 44.20 | 36.12 | 70.39 | 54.76 | 44.77 |
| 0.20 | 56.14 | 48.22 | 32.20 | 70.51 | 59.75 | 39.88 |
| 0.30 | 56.00 | 52.34 | 28.14 | 70.34 | 64.82 | 34.82 |

**Table 3.** Average values of the deviation of the microwave background temperature and its symmetrical and antisymetrical components obtained as a result of the modeling when the symmetry coefficient is changed within the range ( $-0.3 < k_{symm} < 0.3$ ).

In Tables 4 and 5 we present the sums, the differences and their relations for the average absolute values and root mean squares of the symmetrical and antisymmetrical components from the Table 3.

| $k_{symm}$ | $|T^a_{symm}|-|T^a_{asymm}|$ | $|T^a_{symm}|+|T^a_{asymm}|$ | $(|T^a_{symm}|-|T^a_{asymm}|)/(|T^a_{symm}|+|T^a_{asymm}|)$ |
|---|---|---|---|
| -0.30 | -24.30 | 80.43 | -0.302 |
| -0.20 | -15.92 | 80.57 | -0.198 |
| -0.10 | -8.09 | 80.56 | -0.100 |
| -0.08 | -6.67 | 80.58 | -0.083 |
| -0.06 | -4.63 | 80.49 | -0.058 |
| -0.04 | -3.29 | 80.50 | -0.041 |
| -0.02 | -1.82 | 80.40 | -0.023 |
| **0.00** | -0.34 | 80.35 | -0.004 |
| 0.02 | 1.40 | 80.29 | 0.017 |
| 0.04 | 3.21 | 80.44 | 0.040 |
| 0.06 | 4.77 | 80.44 | 0.059 |
| 0.08 | 6.36 | 80.60 | 0.079 |
| 0.10 | 8.08 | 80.33 | 0.101 |
| 0.20 | 16.02 | 80.43 | 0.199 |
| 0.30 | 24.20 | 80.48 | 0.301 |

**Table 4.** Sums, differences and their relations for the average absolute values of the symmetrical and antisymmetrical components from the Table 3.

| $k_{symm}$ | $T^q_{asymm} - T^q_{symm}$ | $T^q_{asymm} + T^q_{asymm}$ | $(T^q_{asymm} - T^q_{asymm})/(T^q_{asymm} + T^q_{asymm})$ |
|---|---|---|---|
| -0.30 | -30.14 | 99.71 | -0.302 |
| -0.20 | -19.75 | 99.78 | -0.198 |
| -0.10 | -10.01 | 99.76 | -0.100 |
| -0.08 | -8.18 | 99.77 | -0.082 |
| -0.06 | -5.74 | 99.57 | -0.058 |
| -0.04 | -4.11 | 99.64 | -0.041 |
| -0.02 | -2.20 | 99.73 | -0.022 |
| **0.00** | -0.42 | 99.55 | -0.004 |
| 0.02 | 1.81 | 99.49 | 0.018 |
| 0.04 | 3.90 | 99.66 | 0.039 |
| 0.06 | 5.88 | 99.68 | 0.059 |
| 0.08 | 7.89 | 99.87 | 0.079 |
| 0.10 | 9.99 | 99.53 | 0.100 |
| 0.20 | 19.87 | 99.63 | 0.199 |
| 0.30 | 30.00 | 99.64 | 0.301 |

**Table 5.** Sums, differences and their relations for the aroot mean squares of the symmetrical and antisymmetrical components from the Table 3.

Computer modeling results presented in the tables 3,4,5 show that the average values of the deviation of the temperature and its symmetrical and antisymmetrical components are related to the symmetry coefficient by the expressions:

$$k_{symm} = (|T^a{}_{symm}|-|T^a{}_{asymm}|)/(|T^a{}_{symm}|+|T^a{}_{asymm}|); \tag{5}$$

$$(|T^a{}_{symm}|+|T^a{}_{asymm}|) = const\,(k_{symm}); \tag{6}$$

$$k_{symm} = (T^q{}_{symm} - T^q{}_{asymm})/(T^q{}_{symm} + T^q{}_{asymm}); \tag{7}$$

$$(T^q{}_{symm} + T^q{}_{asymm}) = const\,(k_{symm}); \tag{8}$$

$$|T^a| = const\,(k_{symm}); \tag{9}$$

$$T^q = const\,(k_{symm}). \tag{10}$$

Inserting the values from the table 1 we obtain the average value of the symmetry coefficient for the whole celestial sphere:

from (5)  $k_{symm} = -3.03/79.39 = -0.038 = -3.8\%$;

from (7)  $k_{symm} = -3.95/99.71 = -0.040 = -4.0\%$.

The inaccuracy conditioned by the inexact calculation of the average values from the data in FITS format (wmap_ilc_7yr_v4.fits), can be evaluated by the magnitude of the calculation value $T^a$ from the Table 1. This value should be equal to zero on account of the calculation of the temperature deviation from the average value. The evaluation in relation of $(T^a / T^q)$ gives the inaccuracy of 1.1% .

By the presented data the averaged for the whole celestial sphere symmetry coefficient $k_{symm} = -4\pm1\%$ (antisymmetry 4%).

The distribution of the central symmetry on the celestial sphere is presented on the Picture 5. On the map we can see the smoothed distribution in Mollweide projection of the symmetry coefficient $k_{symm}$, calculated from the WMAP data by the expression (5). The dark segments are the zones of the negative central symmetry ($k_{symm}<0$), the light ones — of the positive central symmetry ($k_{symm}>0$). The step height of the gray scale is about 0.25. We can see from the distribution that both positive and negative values of $k_{symm}$ on many segments of the celestial sphere with angular measures less than 15-20$^0$, are more than 25%, and reach 50% and more for some of them.

So this way the obtained above average value $k_{symm} = -4 \pm 1\%$ is the result of the simultaneous action of the two described above mechanisms of the formation of the primary microwave background inhomogeneities that provoke the central symmetry of opposite signs. It is natural to assume that both these mechanisms act not on the separate segments but on the whole surface of the celestial sphere and the positive or negative sign of their result on one or another celestial segment is related not to the absence but to the prevalence of one of them. If their resulting value (difference) reaches ±50%, we can expect that the central symmetry coefficients conditioned by each of these mechanisms can be close to 1.

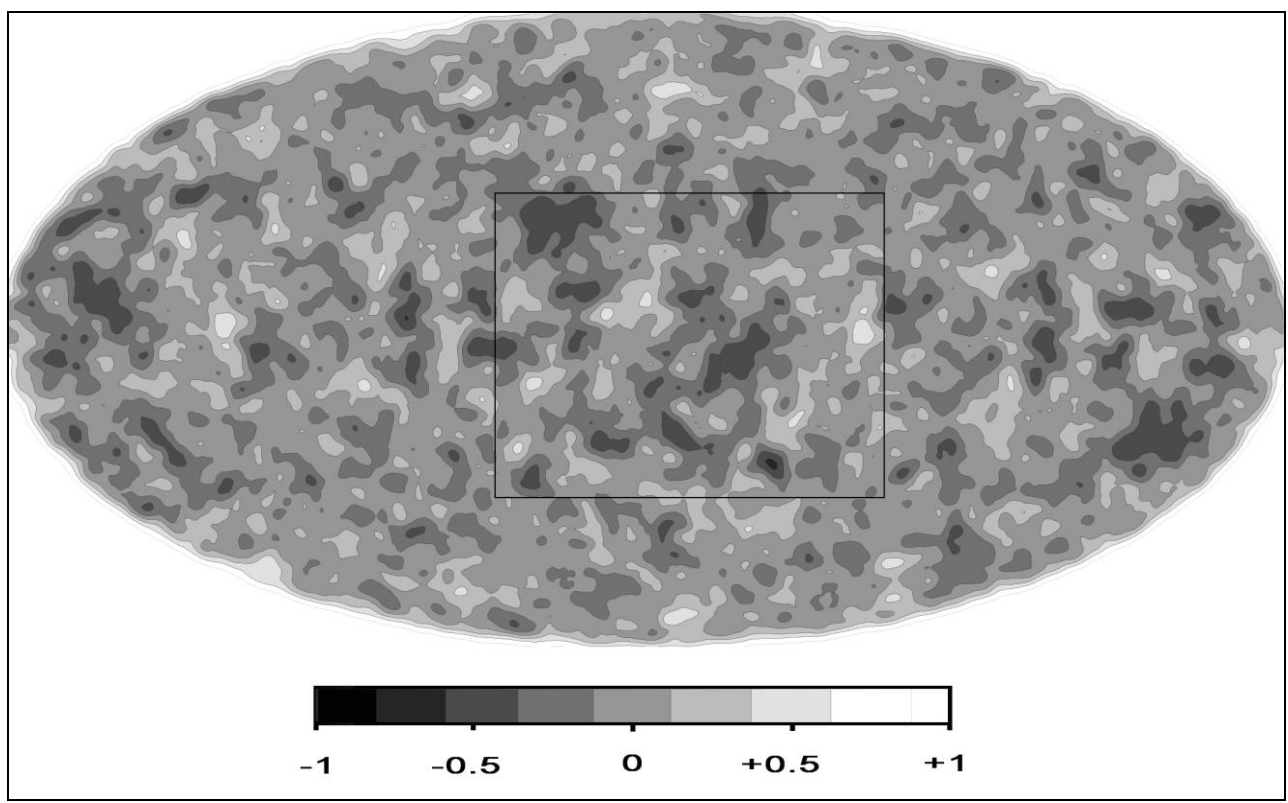

**Picture 5.** Manifestation of the central symmetry of the microwave background on the map in Mollweide projection. We present the smoothed distribution of the central symmetry coefficient, The dark segments correspond to ($k_{symm}<0$) ("antisymmetry"), the light ones — to ($k_{symm}>0$). The rectangle points out the fragment with angular measures about $110^0$ x $70^0$, presented without smoothing on the Picture 6.

On the Picture 6 we present a pointed out on the Picture 5 fragment of the distribution $k_{symm}$, that includes the areas of both positive and negative average values of the symmetry coefficient, without smoothing. On the significantly "dark" and "light" segments the image consists of mainly dark or mainly light elements. The segments that correspond to almost zero average values of $k_{symm}$, include both dark and light elements which we consider a confirmation of the assumption about two opposite to each other mechanisms of the formation of the central symmetry with $k_{symm} \mid \approx 1$.

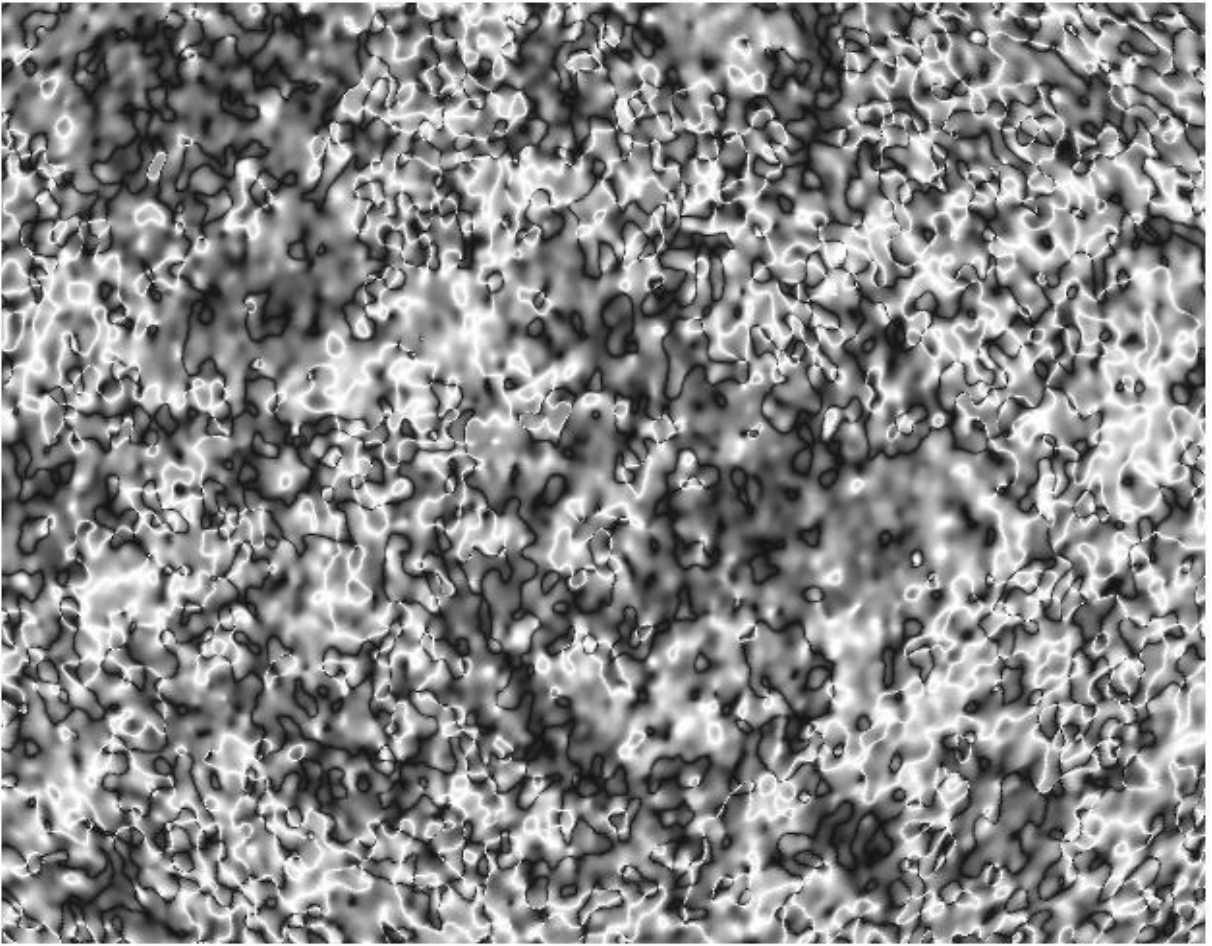
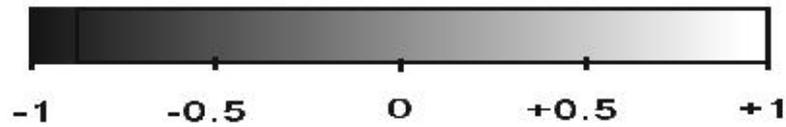

**Picture 6.** Fragment of the distribution of the central symmetry coefficient without smoothing on the surface of the celestial sphere.

A big amount of black and white lines on the picture draws attention. This feature is explained by the fact that $k_{symm}$ is proportional to the difference of the two absolutely different values. To calculate without averaging

$$k_{symm} = (|T_{symm}|-|T_{asymm}|)/(|T_{symm}|+|T_{asymm}|); \qquad (11)$$

whence it follows that white lines are the lines of zero value of $T_{asymm}$, and black lines are the lines of zero value of $T_{symm}$. But from (1),(2) we obtain that a white line corresponds to the equality $T_1 = T_{symm}$, and the black one to the equality $T_1 = T_{asymm}$, i.e. 100% of the central symmetry of the positive or negative sign which also confirms the assumption made before.

### 4. Conclusion

The visual and numerical analysis of the microwave background inhomogeneities performed in this work proves that the distribution of the deviations of the microwave background temperature does not obey the law of the axisymmetry but the law of the central symmetry which manifests as a simultaneous existence of two kinds of symmetry of the opposite sign. The positive symmetry corresponds to the addition to the deviation of the temperature in the opposite (centrally symmetrical) point of the celestial sphere with the same sign than the deviation of the temperature in the observed point, and the negative one – with the opposite sign ("antisymmetry").

After the comparison of the WMAP data analysis results and computer modeling it is established that the symmetry coefficient defined as a relation of the addition to the deviation of the temperature in the opposite point to the initial deviation of the temperature is equal to the relation of the difference of the average absolute values of the symmetrical and antisymmetrical components of the deviation of the temperature to their sum and also the relation of the difference of root mean squares of the symmetrical and antisymmetrical components of the deviation of the temperature to their sum. The calculation of these values and the map of the symmetry coefficient distribution in Mollweide projection testifies on the approximately equal (bipolar) contribution of both components with the prevalence of one of them on some segments of the celestial sphere and the others – on the others. The average value of the symmetry coefficient on the segments of the celestial sphere with angular measures less than $15\text{-}20^0$ changes within the range from -50% to +50%. The mean over the whole celestial sphere symmetry coefficient $k_{symm} = -4\pm1\%$ (antisymmetry 4%).

The small scale structure includes the inhomogeneities with angular measures about $1\text{-}2^0$, separated by the lines corresponding to the equality of the deviation of the temperature and its symmetrical or antisymmetrical component. This particularity can point out the fact that the final distribution of the microwave background can be the result of the simultaneous bipolar action of the mechanisms of the central symmetry and central antisymmetry with the symmetry coefficients close to 100%.

The existence of two simultaneously acting mechanisms of the central symmetry can be related to the influence of the different kinds of the primary inhomogeneities induced by movement of the matter and inhomogeneities of its density at the period of recombination. Inhomogeneities related to the matter movement can induce the antisymmetrical component and the inhomogeneities of its density – the symmetrical one of the observed distribution of the deviation of the temperature.

July 2010

# APPENDIX

To get a pair of comparable images the similar points (the uniformly located ones) of which correspond to the diametrically opposite points of the celestial sphere, the initial image of the right hemisphere should be inverted. To reveal the effect of antisymmetry we will remove from the images the colors that correspond to the large deviations of the opposite sign on the opposite hemispheres. As a result of such a preprocess we obtain 2 pairs of hemispheres images demonstrated on the picture A1.

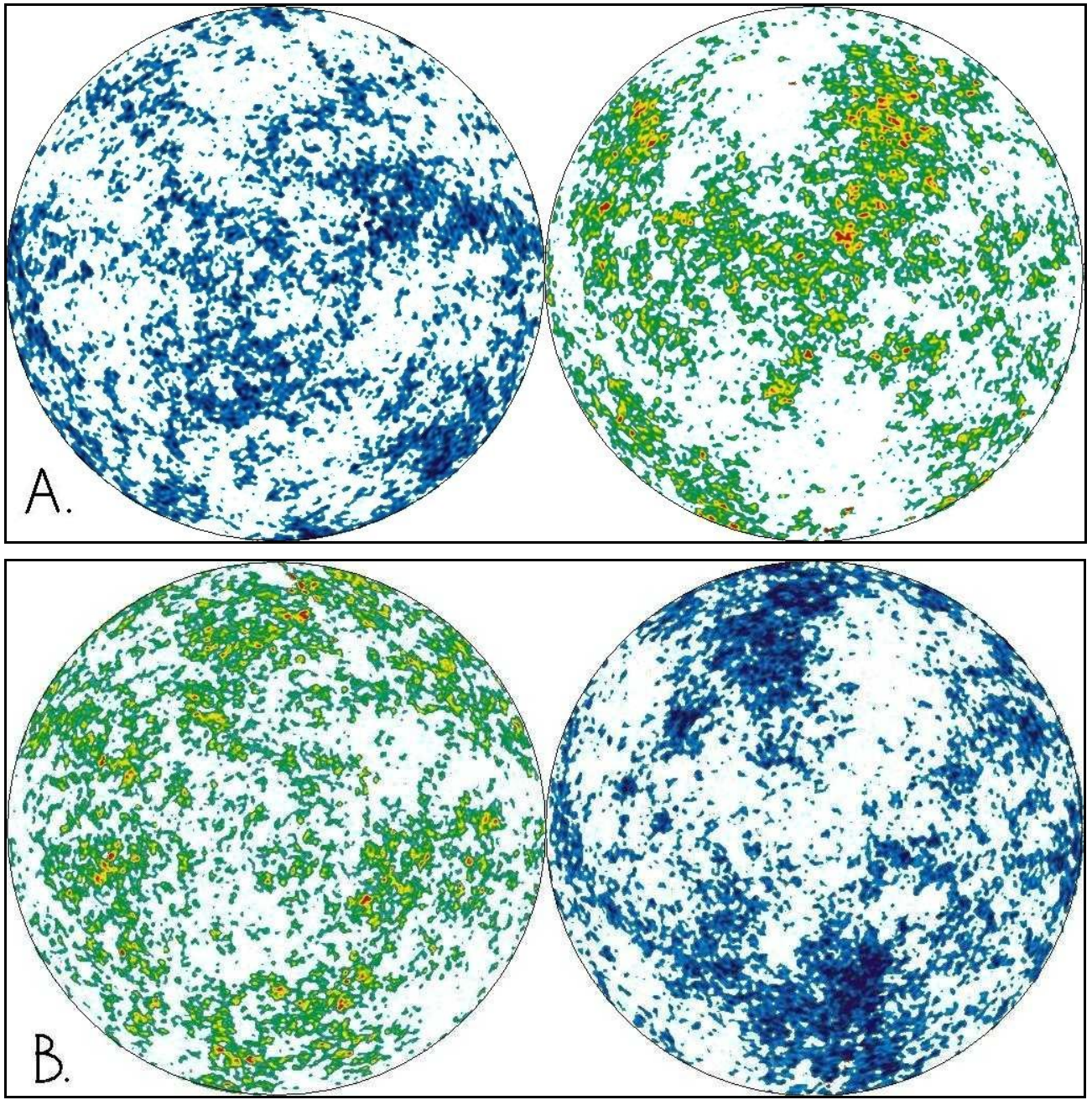

**Picture A1.** The pairs of the images of the celestial hemispheres obtained after processing the initial WMAP images. The pictures of right hemispheres are inverted.
Part A. On the right hemisphere we kept the colors that correspond to the positive deviations of the temperature, on the left – to the negative ones.
Part B. On the left hemisphere we kept the colors that correspond to the positive deviations of the temperature, on the right – to the negative ones.

To make a confident visual analysis of the coinciding elements we can use the ability of the visual analyzer to reveal the reflective symmetry of the examined images. With this objection in mind we distinguished and joined in pairs the uniformly located fragments of the left and right images, moreover the right image is in mirror reflection. On the picture A2 there are the examples of such pairs of the fragments of the images shown on picture A1. What is important here is to emphasize that the left and the right halves of each pair belong to the opposite halves of the celestial sphere and are located there centrosymmetrically.

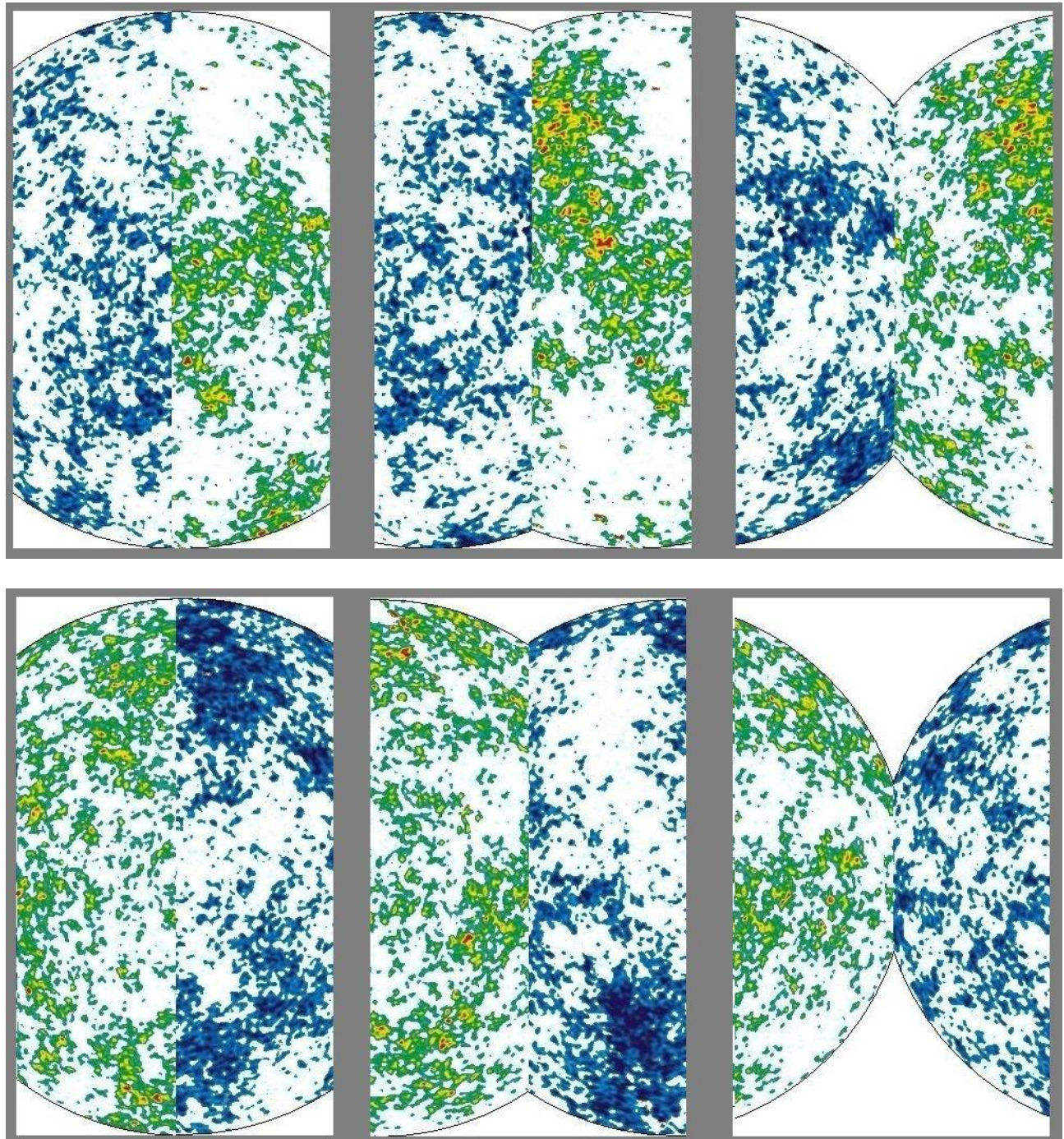

**Picture A2.** Comparison of the fragments of the images of microwave background inhomogeneities that correspond to the diametrically opposite regions of the celestial sphere. The right halves are taken in the mirror reflection.

On the picture A3 the elements of the reflective symmetry that can be distinguished on the pairs of fragments from picture A2 are marked with lines.

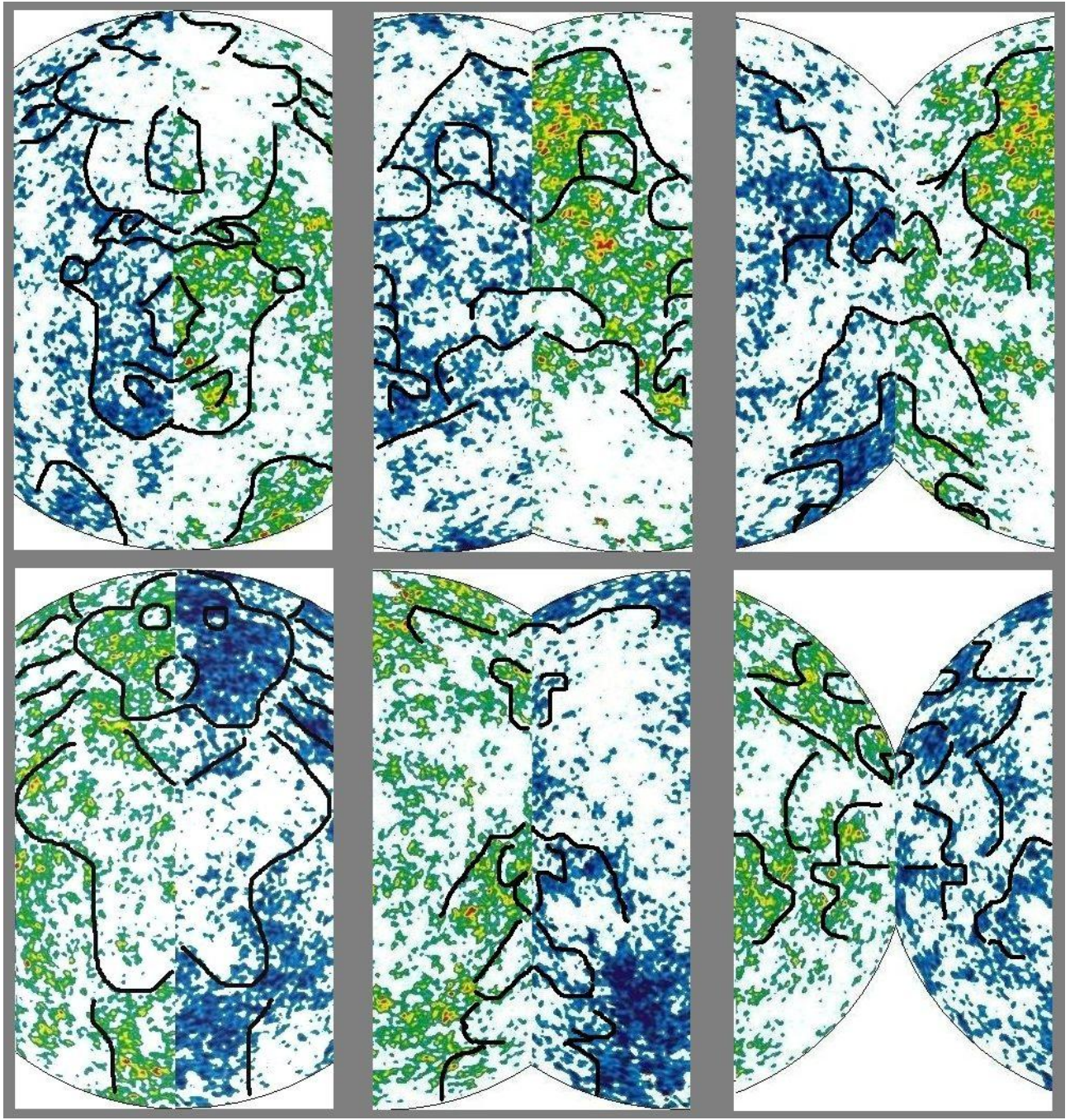

**Picture A3.** On the picture the elements of the reflective symmetry that can be distinguished on the pairs of fragments from picture A2 are marked with lines.

On the picture A4 we construct the images of the hemispheres coinciding to the ones from the picture A1 from the ones on picture A3 and other analogous fragments with the marks applied. To construct the right hemisphere the rights halves of the pairs of fragments were mirror-like reflected again, and after that the marked elements of symmetry took the significance of the structural similarity elements of the right and left images.

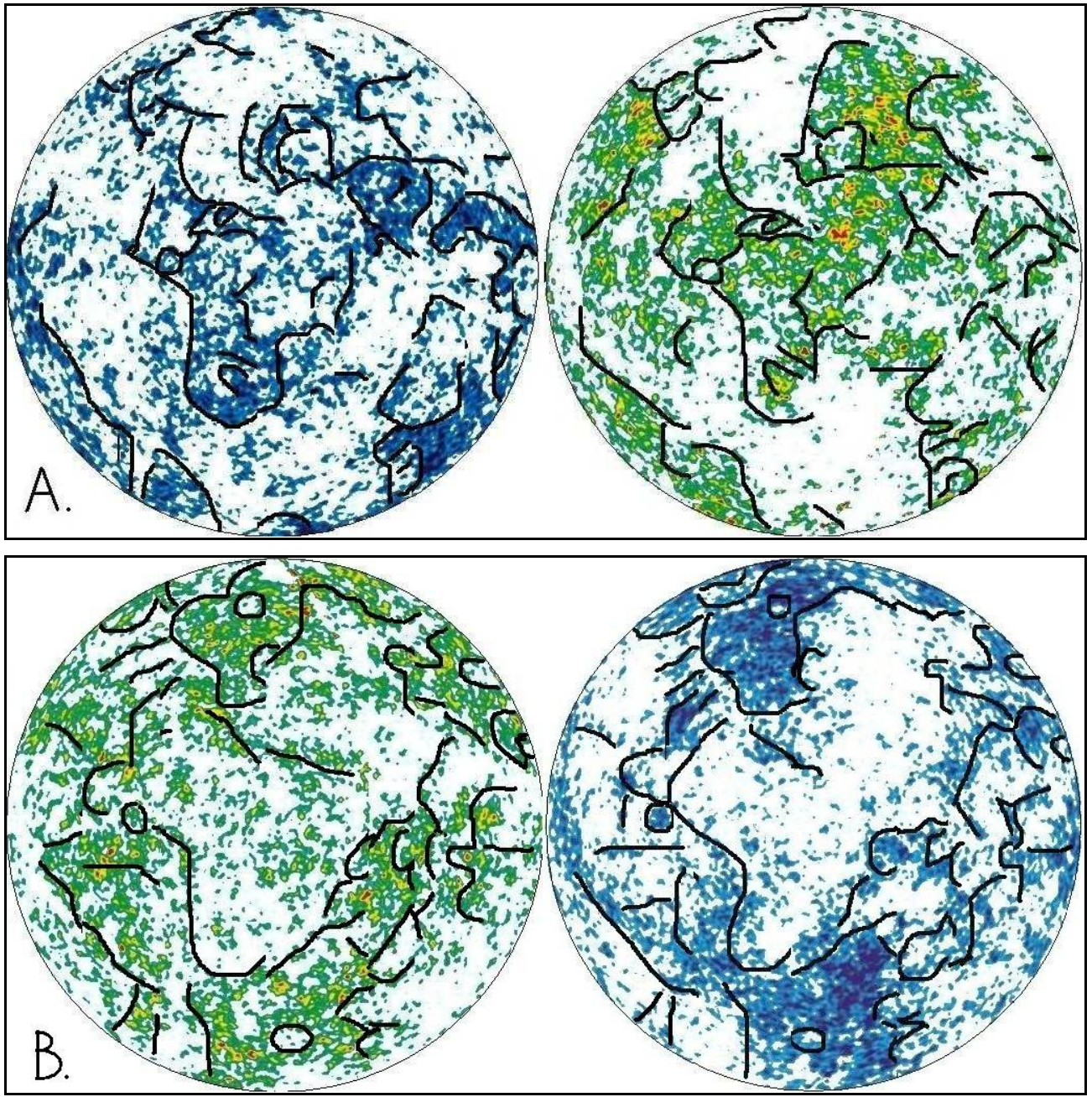

**Picture A4.** Pairs of images of the celestial hemispheres reconstructed from the fragments with marks (Picture A3).

The summarized pattern of these marks on the picture A4 shows the existence of the explicit quantity of the structural similarity elements on the opposite hemispheres including not only the ones of large-scale but also the elements with angular measures $5^0$-$10^0$. Certainly, some symmetry and structural similarity elements on the pictures A3 and A4 can be the result of some accidental resemblance. But the assumption that it can be attributed to all of the marked elements is unlikely, considering their significant quantity.